\documentclass[10pt,conference,letter]{IEEEtran}
\usepackage{float}

\usepackage{longtable}

 \usepackage[usenames,dvipsnames]{pstricks}
 \usepackage{epsfig}
\usepackage{epstopdf}
 \usepackage{pst-grad} 
 \usepackage{pst-plot} 
\usepackage{graphics}

\usepackage{footnote}
\usepackage{cite}
\usepackage{amsmath}
\usepackage{amssymb}
\usepackage{amsfonts}
\usepackage{tikz}

\begin{document}

\title{An Evaluation of Digital Image Forgery Detection Approaches}

\author{\IEEEauthorblockN{Abhishek Kashyap, Rajesh Singh Parmar, Megha Agarwal, Hariom Gupta}
\IEEEauthorblockA{Department of Electronics and Communication Engineering,\\ Jaypee Institute of Information Technology,\\ Noida-201304, Uttar Pradesh, India.\\
Email: abhishek.kashyap@jiit.ac.in, rajesh.parmar@jiit.ac.in, \\ megha.agarwal@jiit.ac.in,  hariom.gupta@jiit.ac.in}
\and
}






\maketitle
\thispagestyle{plain}\pagestyle{plain}

\begin{abstract}

With the headway of the advanced image handling
software and altering tools, a computerized picture can be
effectively controlled. The identification of image manipulation is
vital in light of the fact that an image can be utilized as legitimate
confirmation, in crime scene investigation, and in numerous
different fields. The image forgery detection techniques intend
to confirm the credibility of computerized pictures with no prior
information about the original image. There are numerous routes
for altering a picture, for example, resampling, splicing, and copy-move.
In this paper, we have examined different type of image
forgery and their detection techniques; mainly we focused on
pixel based image forgery detection techniques.

\end{abstract}


\section{Introduction}

Imitations are not new to humanity but rather are an exceptionally old issue. In the past it was restricted to craftsmanship and writing yet did not influence the overall population. These days, because of the headway of computerized picture handling software and altering devices, a picture can be effortlessly controlled and changed \cite{J1}. It is extremely troublesome for people to recognize outwardly

whether the picture is unique or manipulated. There is fast increment in digitally controlled falsifications in standard media and on the Internet \cite{J2}. This pattern shows genuine vulnerabilities and abatements the credibility of digital images. In this manner, creating procedures to check the honesty and realness of the advanced pictures is essential, particularly considering that the pictures are introduced as evidence in a court of law, as news things, as a part of restorative records, or as money related reports. In this sense, image forgery detection is one of the essential objective of image forensics \cite{J3}.

The main objective of this paper is:
To present various aspect of image forgery detection;
To review some late and existing procedures in pixel-based image forgery detection;
To give a comparative study of existing procedures with their advantages and disadvantages.

The rest of the paper is organized as follows. A review of image forgery detection have presented in first section. In second section we discuss different type of digital image forgery. In third section we present digital image forgery detection method. In fourth Section  we introduce and discuss about different existing techniques of pixel-based image forgery detection, mainly copy-move. Comparison of various detection algorithms are given in fifth section and the last section gives the conclusion of this paper.

\section{TYPES OF DIGITAL IMAGE FORGERY}
Picture altering is characterized as "adding, changing, or deleting some important features from an image without leaving any obvious trace” \cite{J2}. There have been different techniques utilized for forging an image. Taking into account the methods used to make forged images, digital image forgery can be isolated into three primary classifications: Copy-Move forgery, Image splicing, and Image resampling.

\subsection{Copy-Move Forgery}

\begin{figure}
        \begin{center}
        \centering
        \includegraphics[height=3.8cm]{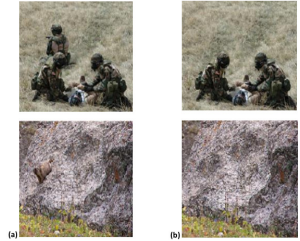}
        \caption{(a) Real image				(b) Forged version \cite{1}}\label{abhi-1}
        \end{center}   		
	 \end{figure}

In copy-move forgery (or cloning), some part of the picture of any size and shape is copied and pasted to another area in the same picture to shroud some important data as demonstrated in Figure \ref{abhi-1}. As the copied part originated from the same image, its essential properties such as noise, color and texture don't change and make the recognition process troublesome.

\subsection{Image Forgery using Splicing}

\begin{figure}
        \begin{center}
        \centering
        \includegraphics[height=4cm]{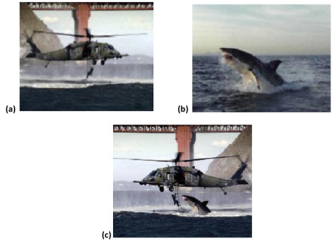}
        \caption{(a) Image (i); (b) Image (ii); (c) Combined image \cite{2}}\label{abhi-2}
        \end{center}   		
	 \end{figure}

Image splicing uses cut-and-paste systems from one or more images to create another fake image. When splicing is performed precisely, the borders between the spliced regions can visually be imperceptible. Splicing, however, disturbs the high order Fourier statistics. These insights can therefore be utilized as a part of distinguishing phony. Figure \ref{abhi-2}, demonstrates a decent sample of image splicing in which the pictures of the shark and the helicopter are merged into one picture.

\subsection{Image Resampling}

\begin{figure}
        \begin{center}
        \centering
        \includegraphics[height=2.3cm]{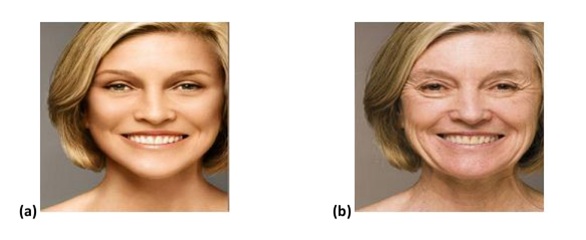}
        \caption{(a) The real image  (b)Result of image retouching \cite{3}}\label{abhi-3}
        \end{center}
    \end{figure}

To make an astounding forged image, some selected regions have to undergo geometric transformations like rotation, scaling, stretching, skewing, flipping and so forth.
The interpolation step plays a important role in the resampling process and introduces non-negligible statistical changes. Resampling introduces specific periodic correlations into the image.
These correlations can be utilized to recognize phony brought about by resampling. In Figure \ref{abhi-3}, the picture on the left is the original image while the one on the right is the forged image obtained by rotation and scaling it.

\section{DIGITAL IMAGE FORGERY DETECTION METHODS}
Digital image forgery detection techniques are grouped into two categories such as active approach and passive approach.  In the active approach, certain information is embedded inside an image during the creation in form of digital watermark. Drawback of this approach is that a watermark must be inserted at the time of recording, which would limit to specially equip digital cameras.
In the passive approach, there is no pre-embedded information inside an image during the creation. This method works purely by analyzing the binary information of an image. Passive image forgery detection techniques roughly grouped into five categories \cite{J4}.

\subsection{ Pixel-based image forgery detection}
 Pixel-based techniques accentuate on the pixels of the digital image. These techniques are generally classified into four sorts such as copy-move, splicing, resampling and statistical. We are concentrating just two sorts of techniques copy-move and splicing in this paper. This is most common image manipulation technique amongst the well-known phony identification techniques.

\subsection{ Format-based image forgery detection}
  Format based techniques are another kind of image forgery detection techniques. These are mainly based on image formats, in which JPEG format is preferable. Statistical correlation introduced by specific lossy compression schemes, which is helpful for image forgery detection. These techniques can be partitioned into three sorts such as JPEG quantization, Double JPEG and JPEG blocking. If the image is compressed then it is exceptionally hard to identify fraud however these techniques can detect forgery in the compressed image.

\subsection{ Camera-based image forgery detection}
  Whenever we take a picture from a digital camera, the picture moves from the camera sensor to the memory and it experiences a progression of processing steps, including quantization, color correlation, gamma correction, white adjusting, filtering, and JPEG compression. These processing steps from capturing to saving the image in the memory may shift on the premise of camera model and camera antiques.

These techniques work on this standard. These methods can be separated into four classes such as chromatic aberration, color filter array, camera response and sensor noise.

\subsection{ Physical environment-based image forgery detection}
  These techniques basically based on three dimensional interactions between physical object, light and the camera. Consider the creation of a forgery showing two movie stars, rumored to be romantically involved, strolling down a nightfall shoreline. Such a picture may be made by grafting together individual pictures of each movie star. In this manner, it is frequently hard to exactly match the lighting effects under which each individual was initially captured.

Contrasts in lighting across an image can be utilized as proof of altering. These techniques work on the basis of the lighting environment under which an article or picture is caught. Lighting is very important factor for capturing an image. These techniques are isolated into three classifications such as light direction (2-D), light direction (3-D) and light environment.

\subsection{ Geometry-based image forgery detection}
  These techniques basically based on principal point i.e. projection of the camera center onto the image plane, that make measurement of the object in the world and their position relative to camera.

Grooves made in firearm barrels confer a twist onto the shot for increased accuracy and range. These grooves acquaint to some degree particular markings to the bullet fired, and can consequently be utilized with a particular handgun. In the same soul, several image forensic techniques have been produced that particularly display relics presented by different phases of the imaging procedure.

Geometry-based image forgery detection methods are separated into two classes such as principle point and metric measurement \cite{J4}.

\section{PIXEL BASED EXISTING IMAGE FORGERY DETECTION TECHNIQUES}

\begin{figure}
        \begin{center}
        \centering
        \includegraphics[width=13cm,height=6.5cm]{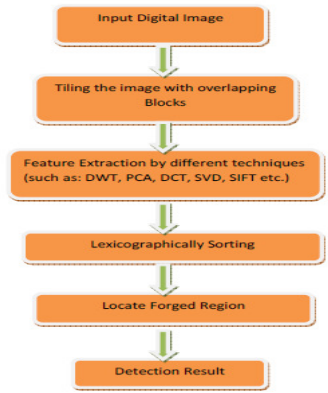}
        \caption{General block diagram of copy-move image forgery detection system.}\label{abhi-4}
        \end{center}
    \end{figure}

There are numerous methodologies that have been proposed by different authors for identifying pixel-based image forgery.
Figure \ref{abhi-4} demonstrates the general procedure of detecting copy-move image forgery \cite{J2}.

PCA: principal component analysis; DCT: discrete cosine transform; DWT: discrete wavelet transform; SVD: singular value decomposition; SIFT: scale invariant feature transform; SURF: speeded up robust features.

Fridrich et al. \cite{J13} proposed a method for identifying copy-move image forgery in 2003. In this method, the image is divided into overlapping blocks (16 x16), and DCT coefficients are used for feature extraction of these blocks. At that point, the DCT coefficients of blocks are lexicographically sorted. After lexicographical sorting, comparable squares are distinguished and forged region are found. In this paper authors perform robust retouching operations in the image. But authors have not performed some other vigor test.

Popescu et al. \cite{J14} proposed a method for identifying duplicate image regions in 2004. In this method, authors applied PCA on fixed-size image of block size (16 x 16, 32 x 32), then computed the Eigen values and eigenvectors of each block. The duplicate regions are automatically detected by using lexicographical sorting.
This algorithm is an efficient and robust technique for image forgery detection even if the image is compressed or noisy.

Kang and Wei \cite{J8} proposed the utilization of SVD to distinguish the altered areas in a digital image in 2008. In this paper Authors utilized SVD for extracting feature vector and dimension reduction. Similar blocks are identified by using lexicographical sorting on row and column vectors and to detect forged regions. This method is robust and efficient.

Lin et al. \cite{J15} proposed quick copy-move forgery detection technique in 2009. In this paper Authors utilized PCA for finding features vectors and dimension reduction after that Radix sort is applied on feature vectors to recognize phony. This algorithm is proficient and functions admirably in noisy and compressed images.

Huang et al. \cite{J9} proposed copy-move forgery detection in digital images using SIFT algorithm in 2009. In this paper, authors presented SIFT calculation algorithm using feature matching. This algorithm gives great results even when picture is compressed or noisy.

Li et al. \cite{J10} proposed a copy-move forgery detection based on sorted neighborhood approach by using DWT and SVD in 2007. In this paper, authors utilized DWT and disintegrated into four sub-groups. SVD was utilized in low-frequency sub-bands for dimension reduction. At that point, they connected lexicographical sorting on particular quality vector and the forged region is recognized. They tried this algorithm for gray-scale and colour images. This algorithm is robust.

Luo et al. \cite{J16} proposed a strong identification of duplicated region in digital images in 2006. In this paper, authors divide an image into overlapping blocks and then apply similarity matching algorithm on these blocks. The similarity matching algorithm recognizes the copy –move forgery in the given image. This method additionally meets expectations in the JPEG compression, additive noise and Gaussian blurring.

Zhang et al. \cite{J17} proposed a new method for cop-move forgery detection in digital image in 2008. Authors utilized DWT and divide given image into four non-overlapping sub-images and phase correlation is adopted to compute the spatial offset between the copy-move forgery regions. At that point, they applied similarity matching algorithm between the pixels for detecting forged regions. This method functions admirably in the highly compressed image and extremely effective with lower computational time as compared with other methods.

Kang et al. \cite{J18} proposed a method to detect copy-move forgery in digital image in 2010. In which firstly image is divided in sub-blocks then applied improved SVD on each blocks. At that point, similarity matching is performed on each blocks based on the lexicographically sorted SV vectors. Finally image forgery region is detected.

Ghorbani et al. \cite{J11} proposed a method to detect copy-move forgery based on DWT-DCT (QCD) in 2011. Authors utilized DWT to divide image into sub-bands, then performed DCT-QCD (Quantization coefficient decomposition) in row vectors to reduce vector length. Shift vector is computed after lexicographically sorting of the row vectors, then it is compared with threshold and finally duplicated region of an image is highlighted.

Lin et al. \cite{J7} proposed an integrated method for copy-move and splicing forgery detection in 2011. To begin with, the authors changed over a picture into the YCbCr colour space. For copy-move detection, SURF is used. For splicing detection, image is firstly divided into sub-blocks, then DCT is applied for feature extraction in each blocks. This method works well in both copy-move and splicing forgery detection.

Qu et al. \cite{J6} proposed a algorithm to detect splicing image forgery with visual cues in 2009. Authors used a detection window and divided it into nine sub-squares. VAM (visual consideration model) is used to distinguish an obsession point and afterward feature extraction  is used to extract the spliced region in the digital image.

Lin et al. \cite{J19} proposed an automatic and quick altered JPEG image detection technique using analysis of DCT coefficient in 2009. Authors have utilized DCT coefficient and Bayesian approach for feature extraction, then similarity matching algorithm is used to detect duplicated region map.

Huang et al. \cite{J22} proposed a method to detect copy-move forgery based on Improved DCT of an image in 2011. In this paper, DCT coefficients are used for finding feature vector. After that similarity matching algorithm is used to identify imitation areas of an image.

Cao et al. \cite{J23} proposed a robust algorithm to detect copy-move forgery in digital image in 2012. In this paper, authors have used DCT for finding DCT coefficients of each block that are represented by circle block and extract feature from each circle block, then searching operation is performed to find similar block pairs for duplicated region map.

 G. Muhammad \cite{1} proposed a blind copy move image forgery detection method using dyadic wavelet transform (DyWT). DyWT is shift invariant and hence more relevant than discrete wavelet transform (DWT) for data analysis.
  In this method First we decompose the input image into approximation (LL1) and detail (HH1) subbands. Then we divide LL1 and HH1 subbands into overlapping blocks and measure the similarity between blocks.
 The main idea is that the similarity between the copied and moved blocks from the LL1 subband should be high, while the one from the HH1 subband should be low due to noise inconsistency in the moved block.
	This method is not relevant for color information instead of converting the color images to gray images. This method is highly efficient method

N. Muhammad \cite{2} proposed a method to detect Copy-move forgery, which is one type of tempering that is commonly used for manipulating the digital images. In this
method a part of an image is copied and is pasted on another region of the image. In this paper efficient non-intrusive method for copy-move forgery detection is explained. This method is based on image segmentation and similarity detection using dyadic wavelet transform (DyWT). Copied and pasted regions are structurally similar and this structural similarity is detected using DyWT and statistical measures. The results show that this method outperforms the stat-of-the art methods. In this paper algorithm effectively detect tempering on the image and no need of the knowledge about any camera and large number of image for decision making. the algorithm can be used for complicated background and texture.

\begin {table*}[]
\caption {Comparative study of existing copy-move forgery detection methods} \label{}
\centering
\begin{tabular}{|p{.2in}|p{1.5in}|p{1.0in}|p{1.3in}|p{1.8in}|p{.5in}|} \hline
\textbf{S.No.} & \textbf{Paper title} & \textbf{Method used} & \textbf{Tampering detection type} & \textbf{Pros/cons}& \textbf{Publication year}\\ \hline

1 & Detection of copy-move forgery in digital image \cite{J13} & DCT & Copy-move region is
detected
& Will not work in noisy
image
& 2003
\\ \hline
2 & Exposing digital forgeries by detecting duplicated image
regions \cite{J14}
& PCA & Exact copy-move
region is detected
automatically
& Time complexity is
high
& 2004
\\ \hline

3 & Robust detection of region duplication in digital image \cite{J16} & Similarity matching & Copy-move region
detected in noisy
conditions
& Time complexity is
reduced \cite{J14}
& 2006
\\ \hline
4 & A sorted neighbourhood approach for detecting duplicate
reason based on DWTand SVD \cite{J10}
& DWT-SVD & Efficiently detects
forged region
& Time complexity is
less compared to
other algorithms
\cite{J14}
& 2007
\\ \hline
5 & A new approach for detecting copy-move forgery detection
in digital image \cite{J17}
& DWT & Exact copy-move
region is detected
& Works well in noisy
and compressed
image
& 2008
\\ \hline
6 &  Detection of copy-move forgery in digital images using
SIFT algorithm \cite{J9}
& SIFT & Copy-move region is
detected
& Detects false result
also
& 2008
\\ \hline
7 & Identifying tampered regions using singular value
decomposition in Digital image forensics \cite{J8}
& SVD & Copy-Move region is
detected accurately
& Will not work in
highly noised and
compressed image
&2008
\\ \hline
8 & Fast copy-move forgery detection \cite{J15} & Improved PCA & Exact Copy-Move
region is detected
& Works well in noisy,
compressed image
& 2009
\\ \hline
9 & Detect digital image splicing with visual cues \cite{J6} & DW-VAM & In spliced image,
forged region is
detected
& Work only in the
Splicing
& 2009
\\ \hline

10 & Fast, automatic and fine-grained tempered JPEG image
detection via DCT coefficient analysis \cite{J19}
& Double Quantization
DCT
& Tampered region is
detected
accurately
& Works only in JPEG
Format
& 2009
\\ \hline

11 & Copy-move forgery detection in digital image \cite{J18} & SVD & Forged region is
detected
& Will not work well in
noisy image
& 2010
\\ \hline

12 & Blind copy move image forgery detection
Using dyadic undecimated wavelet transform \cite{1}	
 & Dyadic undecimated wavelet transform 	& Copy-move region is detected &	Will not work in noisy image & 2011
  \\ \hline

 13 & Copy-move forgery detection using dyadic wavelet transform \cite{2}	&	Dyadic wavelet transform	& Image segmentation and similarity detection	& Not efficient for complicated background and texture	 & 2011
  \\ \hline


  14 & 	Detecting copy-move forgery using non-negative matrix factorization	\cite{3}	& Non-negative matrix factorization (nmf)	& Copy-move region is detected	& Some geometric distortions (e.g. rotation,
Reflection etc.) Can render the method invalid &	2011
  \\ \hline

  15	& Detecting copy-paste forgeries using transform-invariant features \cite{4}		& Transform-invariant features	& Copy-paste forgery detection	& Difficult detection in case of blurred image	& 2011
  \\ \hline

%

  16	& Detection of copy-create image forgery using
Luminance level techniques	\cite{5}	& Luminance level techniques	& Copy-create image forgery	& Time consuming and less accurate	& 2011
  \\ \hline

  17 &	Image copy-move forgery detection
Based on “crossing shadow” division	\cite{6}	& Dwt and crossing shadow	& Copy- move region detected	& Algorithm has advantages of low
Computational complexity	& 2011
\\ \hline
18 &	A fast image copy-move forgery detection method using phase correlation \cite{7}	&	Phase correlation	& Copy –move region detected	 & method is valid in detecting the image region
Duplication and quite robust to additive noise and blurring	& 2012
\\ \hline
19 & 	An evaluation of popular copy-move
Forgery detection approaches	\cite{8}	& DCT, DWT, KPCA, PCA	& Copy –move region detected	 & low computational load and good performance & 	2012
\\ \hline
20	& Copy-move forgery detection based on PHT \cite{9}	&	Polar harmonic transform(PHT)	 & detect the tampered
Regions when they are rotated before being pasted	Scheme can detect the copy-move forgery
When the copied region is rotated before being pasted. & Scheme is not efficient for scaling, local bending in images	& 2012
\\ \hline

21 & Copy-Move Forgery Detection In Digital Images
Based On Local Dimension Estimation	\cite{10}	
&
Local Dimension Estimation	
& Copy-Move Region Detected	
& Less  Computational  Efficiency	
& 2012
\\ \hline
22	& Copy-move image forgery detection using multi-resolution weber
Descriptors	\cite{11}	& Multi-resolution weber
Descriptors	& Copy move region  detected
	& Multi-resolution Weber law descriptors (WLD) extracts
The features from chrominance components, which can give
More information that the human eyes cannot notice.
WLD is a robust image texture
Descriptor and with its extension to different scales
and highly accurate &	2012
\\ \hline
23 &	
 Detection Of Copy-Move Forgery Image Using Gabor Descriptor \cite{12}		
 & Gabor Descriptor &	Copy Move Region Detected &	Highly Accurate And Reliable	& 2012

\\ \hline

\end{tabular}
\end {table*}

\begin {table*}[]
\centering
\begin{tabular}{|p{.2in}|p{1.5in}|p{1.0in}|p{1.3in}|p{1.8in}|p{.5in}|} \hline

24	& Detection of copy-move forgery in digital images
Using radon transformation and phase correlation	\cite{13}	& Radon transformation and phase correlation	& Exact copy move region is detected	& Detect exact forgery even if the
Forged images were underwent some image processing operations
Such as rotation and gaussian noise addition &	2012
\\ \hline
25 &	A robust image copy-move forgery detection
Based on mixed moments	\cite{14}	& Mixed moment, gaussian pyramid transform	& Tampered region is precisely detected	& Accuracy is improved, time complexity and robust features are also solved and algorithm has  some limitations on the smaller tamper
Regions &	2013
\\ \hline
26	& A fast DCT based method for copy move forgery
Detection	\cite{15}	& DCT &	Copy-move region is detected	& Will not work in noisy image	& 2013
\\ \hline
27	& Copy move forgery detection using DWT and
SIFT features \cite{16}	
	& DWT and
SIFT	& Copy move region is detected	& Defects false results also	& 2013
\\ \hline
28 &	Copy move image forgery detection method using
Steerable pyramid transform and texture descriptor \cite{17}	
	& Steerable pyramid transform local binary pattern (LBP)., and texture descriptor	
& Copy move region is detected	& Accuracy is high &	2013
\\ \hline
29 &	Copy move image forgery detection using mutual
Information	\cite{18}	& Mutual information	& Copy-move region detected	& Less accurate	& 2013
\\ \hline
30 &	Copy-move forgery detection in images via
 2D-fourier transform \cite{19}	&	2D- fourier transform	& Copy-move region detected accurately	&  This work
Detects multiple copy move forgery and it also robust to jpeg
Compression attacks even if the quality factor is lower than 50 hence highly accurate
	& 2013
\\ \hline
\newpage
31 &	Copy-move image forgery detection using local binary pattern and
Neighborhood clustering \cite{20}	&	Local binary pattern and
Neighborhood clustering	& Copy-move region detected	&Highly accurate &	2013
\\ \hline
32	& Detection of copy-move forgery using wavelet
Decomposition \cite{21}	&	Wavelet	& Copy-move region detected	& Accuracy is high &	2013
\\ \hline
33 &	Detection of copy-move forgery using krawtchouk moment \cite{22}	
& Krawtchouk moment &	Copy-move region detected	& Works well if the image is noisy or blurred &	2013
\\ \hline
34	& Video copy-move forgery detection and
Localization based on tamura texture features	\cite{23}	& Tamura texture features &
	Copy- move region detected 	& Precision of this
Method is  99.96\%. Hence highly accurate method	& 2013
\\ \hline
35	& A copy-move image forgery detection based on
Speeded up robust feature transform and wavelet
Transforms	\cite{24}	& Speeded up robust feature transform and wavelet
Transforms	& Forged region is detected accurately	& Works well for copy -move	& 2014
\\ \hline
36	& A scheme for copy-move forgery detection in
Digital images based on 2D-DWT \cite{25}	& 	2D-DWT	& Copy- move region is detected	& Works well in noisy and compressed image	& 2014
\\ \hline
37 &  Adaptive Matching For Copy-Move Forgery Detection \cite{26}		
 & Block-Based Methods,
 It Is Proposed To Employ An Adaptive Threshold In The Matching Phase In Order To Overcome This Forgery Problem &	Copy-Move Region Is Detected &	Accuracy Is Less	&
2014
\\ \hline
38	& Copy-move forgery detection based on patch match	\cite{27}	& Patch match, an iterative
Randomized algorithm for nearest-neighbor search	& Copy-move region detected	& Accuracy is high	& 2014
\\ \hline
39	& Copy-move image forgery detection
Based on sift descriptors
And SVD-matching \cite{28}	&	SIFT descriptors
And SVD-matching &	Forged region is detected &	Less efficient in noisy image &	2014
\\ \hline
40 &	Copy-rotate-move forgery detection based on
Spatial domain	\cite{29}	& Spatial domain	& Forged region is detected	& Highly efficient	& 2014
\\ \hline

41	& Copy-rotation-move forgery detection using the
Mrogh descriptor \cite{30}	& 	Mrogh descriptor	& Copy –move region detected	& Highly efficient	&
2014
\\ \hline
42	& Jpeg copy paste forgery detection using bag
Optimized for complex images	\cite{31}	& Bag
Optimized for complex images	& Forged region is detected	& Highly efficient &	2014
\\ \hline
43	& Shape based copy move forgery detection using level set approach \cite{32}	&	Level set approach	& Copy- move region detected	& Time complexity is minimum	& 2014
\\ \hline
44	& Speeding-up sift based copy move forgery
Detection using level set approach	\cite{33}	& SIFT & 	Copy –move region is detected & 	Less efficient	& 2014
\\ \hline
45	& Video frame copy-move forgery detection based
On cellular automata and local binary patterns	\cite{34}	& Cellular automata and local binary patterns	& Copy-move region detected	& Highly efficient	& 2014
  \\ \hline
46	& Detection of splicing forgery using wavelet decomposition \cite{35}		& wavelet decomposition	& Splicing type of forgery detected	& Highly efficient	& 2015
  \\ \hline

\end{tabular}
\end {table*}


Copy-move is a common manipulation in digital images. H. Yao \cite{3} proposed an efficient copy-move detecting scheme with the capacity of some post-processing resistances. The image is divided into fixed-size overlapped blocks, and then non-negative matrix factorization (NMF) coefficients are extracted from list of all blocks. We use lexicographical sorting method to reduce the probability of invalid matching. By measuring the hamming distance of each block pair in the matching procedure, if the distance is shorter than a threshold, we declare  them as the tampering region.


 copy paste forgery is the most common type of image forgery wherein a region from an image is replaced with another region from the same image. P. Kakar \cite{4} proposed a good technique based on transform invariant features. These are basically depend on the Trace transform and achieved by modifying
the MPEG-7 image signature tools descriptors in many aspects. As a result this is highly efficient scheme for image forgery detection.


 L. Li  \cite{9} proposed a best approach for detecting copy-move forgery with rotation. To extract the features of the circular blocks, which are then used to perform block matching polar Harmonic Transform can be used. This method is valid for noisy and rotated figures.


M. Hussain \cite{11} proposed a method to detect copy-move forgery based on Multi-resolution Weber law descriptors
(WLD). The proposed multi-resolution WLD extracts the features from chrominance components, which can give more information that the human eyes cannot notice.Acccuracy rate of the proposed method, can reach   up to 91\% with multi resolution WLD descriptor on the chrominance space of the image.


H. C. Nguyen \cite{13} proposed a method based on non block-matching to detect image copy-move forgery. In this paper  exploiting phase correlation are used. Results of experiments indicate that the method is valid in detecting the image region duplication and quite robust to additive noise and blurring.


S. Kumar \cite{15} proposed a method to detect copy-move forgery. In this method discrete cosine transform(DCT) is used to represent the features of the overlapping blocks.
In the image data set it has detected forgery with good success. Against added Gaussian noise, JPEG compression and small amount of scaling and rotation also, it has shown robustness. However, robustness against more post processing operations like flipping, shearing and local intensity variations may be extended in this algorithm.


M. F. Hashmi \cite{16} proposed a method to detect copy-move forgery using DWT and SIFT. This paper proposed a algorithm of image-tamper detection based on the Discrete Wavelet Transform i.e. DWT. DWT is used for
Dimension reduction, which in turn improves the accuracy of results. First DWT is applied on a given image to decompose it into four parts LL, LH, HL, and HH. Since LL part contains
most of the information, SIFT is applied on LL part only to extract the key features and find descriptor vector of these key features and then find similarities between various descriptor
vectors to conclude that the given image is forged. This method allows us to detect whether image forgery has occurred or not.


       L. Yu \cite{30} proposed a method to detect copy-rotation-move forgery detection using the MROGH descriptor. This paper  discuss a approach, in which screened Harris Corner Detector and the MROGH descriptor are used to gain better feature coverage and robustness against rotation. it is highly efficient method.

\section{COMPARATIVE RESULTS \& DISCUSSION}
 We have discussed various methods that are proposed by various authors to detect image forgery.  The thought process of the considerable number of strategies is to recognize the imitation in the picture yet the procedures are diverse. Table 1 shows the comparison of various copy-move forgery detection methods, which have discussed in this paper.

\begin{figure}
\centering
  \includegraphics[width=3.8in, height=2.3in]{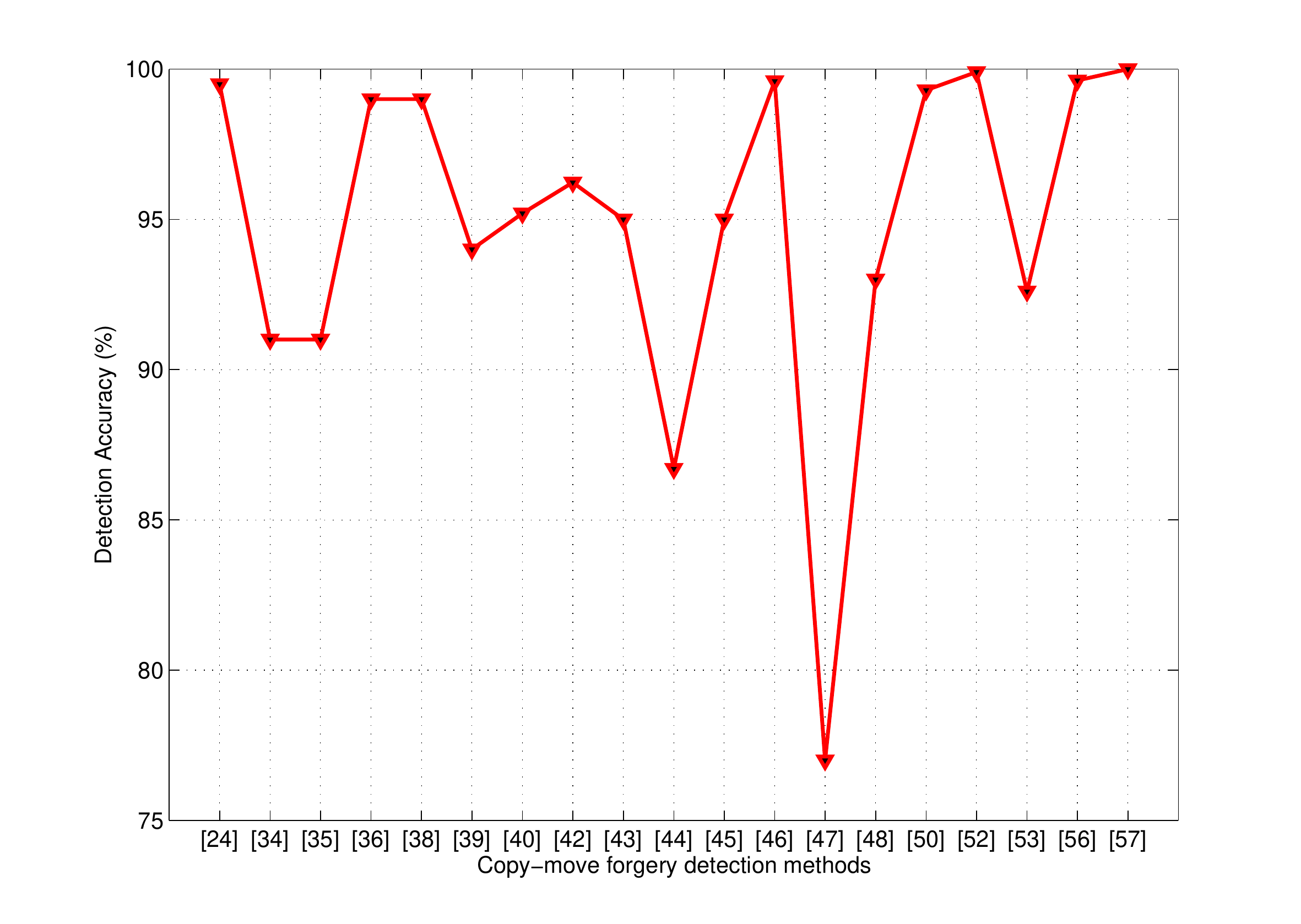}
  \caption{Performance analysis of copy-move forgery detection methods in terms of accuracy}\label{abhi-5}
\end{figure}

\begin{figure}
\centering
  \includegraphics[width=4.0in, height=2.3in]{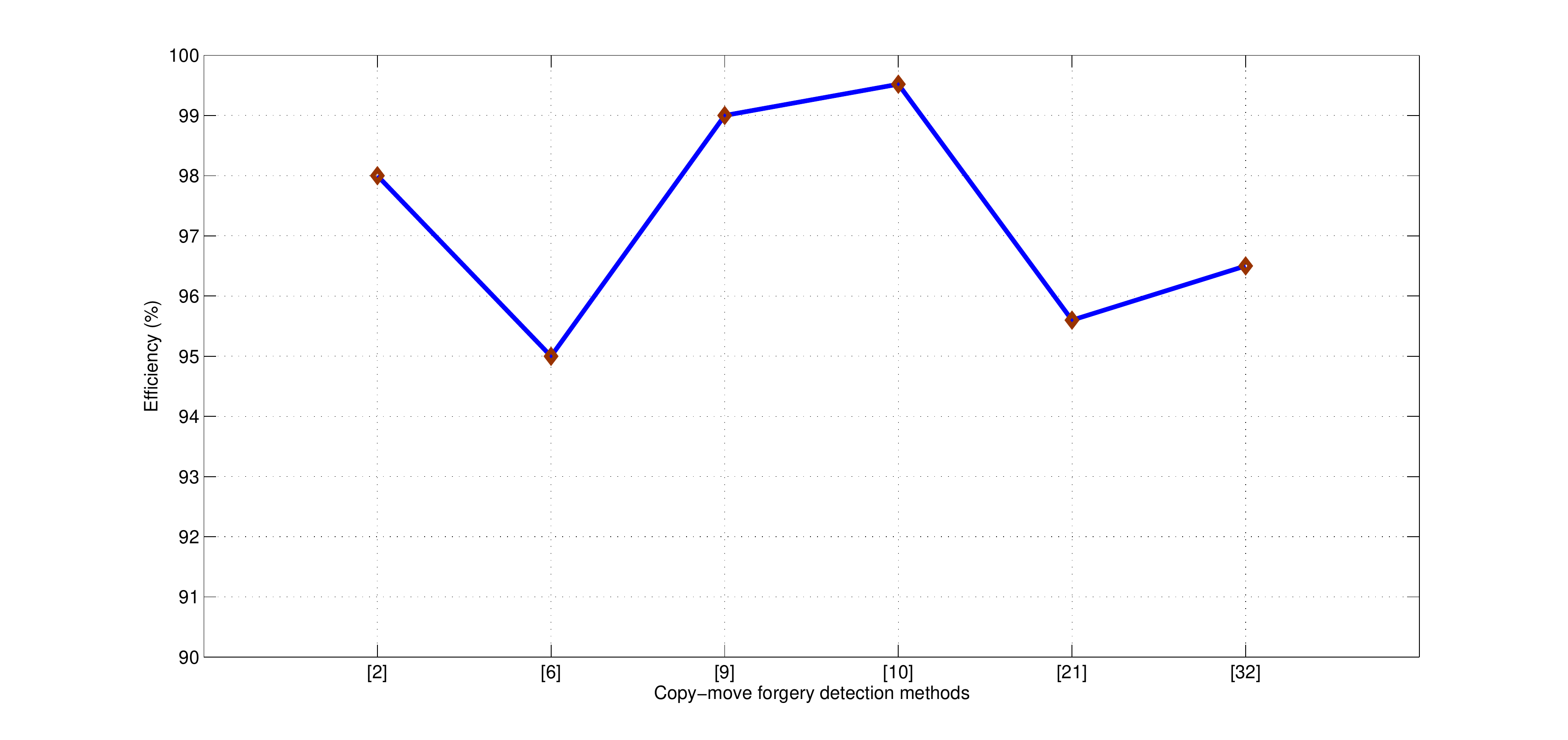}
  \caption{Performance analysis of copy-move forgery detection methods in terms of efficiency}\label{abhi-6}
\end{figure}

Performance analysis of proposed methods \cite{1},\cite{11},\cite{12},\cite{13},\cite{15},\cite{16},\cite{17},\cite{19},\cite{20},\cite{21},\cite{22},\cite{23},\cite{24},\cite{25},\cite{27},\cite{29},\cite{30},\cite{33} and \cite{34} is shown in Figure \ref{abhi-5}, which have detection accuracy 99.5\%, 91\%, 91\%, 99\%, 99\%, 94\%, 95.2\%, 96.23\%, 95\%, 86.7\%, 95\%, 99.6\%, 77\%, 93\%, 99.3\%, 99.9\%, 92.6\%, 99.62\%, and 100\% respectively. Figure \ref{abhi-6} shows  performance analysis of proposed methods \cite{2},\cite{6},\cite{9},\cite{10},\cite{21} and \cite{32}, which have efficiency 98\%, 95\%, 99\%, 99.52\%, 95.60\% and 96.50\% respectively.


\section{CONCLUSION}
In this paper different methodologies of image forgery detection have been surveyed and discussed about. All the approaches and methodologies talked about in this paper have the capacity to recognize fraud. In any case, a few algorithms are not viable regarding identifying actual forged region. On the other hand some algorithms have a time complexity problem. So, there is a need to develop an effective (efficient) and accurate image forgery detection algorithm.

\nocite{*}
\bibliographystyle{IEEE}

\begin{thebibliography}{}

\bibitem{J1}
J. A. Redi, W. Taktak, and J.-L. Dugelay, {\textquotedblleft Digital image forensics:
A booklet for beginners,\textquotedblright}
{\it Multimedia Tool Appl.}, Vol. 51, no. 1, pp. 133-62, Jan. 2011.

\bibitem{J2}
J. Wang, G. Liu, Z. Zhang, Y. Dai, and Z. Wang, {\textquotedblleft Fast and
robust forensics for image region-duplication forgery,\textquotedblright}
{\it Acta
Automatica Sinica}, Vol. 35, no. 12, pp. 1488-95, Dec. 2009.

\bibitem{J3}
V. Tyagi, {\textquotedblleft Detection of forgery in images stored in digital form,\textquotedblright}
{\it Project report submitted to DRDO}, New Delhi, 2010.

\bibitem{J4}
H. Farid, {\textquotedblleft A survey of image forgery detection,\textquotedblright}
{\it IEEE Signal
Process. Mag.},  Vol. 26, no. 2, pp. 16-25, Mar. 2009.

\bibitem{J5}
R. E. J. Granty, T. S. Aditya, and S. Madhu, {\textquotedblleft Survey on passive
methods of image tampering detection,\textquotedblright}
{\it IEEE International
Conference on Communication and Computational Intelligence}, in
(INCOCCI),  2010, pp. 431-436.

\bibitem{J6}
Z. Qu, and G. Qiu, {\textquotedblleft Detect digital image splicing with visual
cues,\textquotedblright}
{\it Lect. Notes Comput. Sci.}, Vol. 5806, pp. 247-326, Jan.
2009.

\bibitem{J7}
S. D. Lin et al., {\textquotedblleft An integrated technique for splicing and copymove
forgery image detection,\textquotedblright}
{\it IEEE 4th International Congress
on Image and Signal Processing (CISP)}, 2011, Vol. 2, pp.
1086-1090.

\bibitem{J8}
X. Kang, and S. Wei, {\textquotedblleft Identifying tampered regions using singular
value decomposition in digital image forensics,\textquotedblright}
{\it International
Conference on Computer Science and Software
Engineering}, 2008, Vol. 3, pp. 926-930.

\bibitem{J9}
H. Huang, W. Guo, and Y. Zhang, {\textquotedblleft Detection of copy-move
forgery in digital images using SIFT algorithm,\textquotedblright}
{\it Pacific-Asia
Workshop on Computational Intelligence and Industrial Application}, Vol. 2, pp. 272-6, Dec. 2008.


\bibitem{J10}
G. H. Li, Q. Wu, D. Tu, and S. J. Sun,  {\textquotedblleft A sorted neighborhood
approach for detecting duplicated regions in image forgeries
based on DWT and SVD,\textquotedblright}
{\it Proceedings of IEEE International
Conference on Multimedia and Expo}, Beijing, Jul. 2007, pp.
1750-1753.


\bibitem{J11}
M. Ghorbani, M. Firouzmand, and A. Faraahi, {\textquotedblleft DWT-DCT
(QCD) based copy-move image forgery detection,\textquotedblright}
{\it 18th
IEEE International Conference on Systems, Signals and Image
Processing (IWSSIP)}, 2011, pp. 1-4.

\bibitem{J12}
I. Amerini et al., {\textquotedblleft A SIFT-based forensic method for copy-move
attack detection and transformation recovery,\textquotedblright}
{\it IEEE Trans. Inf.
Foren. Sec.}, Vol. 6, no 3, pp. 1099-1111, 2011.

\bibitem{J13}
 J. Fridrich, D. Soukal, and J. Lukas, {\textquotedblleft Detection of copy move
forgery in digital images,\textquotedblright}
{\it Proceedings of the Digital Forensic
Research Workshop}, pp. 5-8, Aug. 2003.

\bibitem{J14}
A. C. Popescu, and H. Farid, {\textquotedblleft Exposing digital forgeries by
detecting duplicated image regions,\textquotedblright}
{\it Dept. Comput. Sci., Dartmouth
College, Tech. Rep.}, TR2004-515, 2004. .

\bibitem{J15}
H.-J. Lin, C.-W. Wang, and Y.-T. Kao, {\textquotedblleft Fast copy-move forgery
detection,\textquotedblright}
{\it WSEAS Transaction on Signal Processing},
pp. 188-197,  2009.

\bibitem{J16}
W. Q. Luo, J. W. Huang, and G. P. Qiu, {\textquotedblleft Robust detection of
region-duplication forgery in digital image,\textquotedblright}
{\it 18th International Conference on Pattern Recognition (ICPR)}, 2006, Vol. 4, pp. 746-749.


\bibitem{J17}
J. Zhang, Z. Feng, and Y. Su, {\textquotedblleft A new approach for detecting
copy-move forgery in digital images,\textquotedblright}
{\it IEEE International
Conference on Communication Systems}, China, 2008, pp.
362-366.

\bibitem{J18}
L. Kang, and X.-P. Cheng, {\textquotedblleft Copy-move forgery detection in
digital image,\textquotedblright}
{\it 3rd IEEE International Congress on Image and Signal
Processing (CISP 2010)}, 2010, pp.
2419-2421.

\bibitem{J19}
Z. Lin et al., {\textquotedblleft Fast, automatic and fine-grained tampered JPEG
image detection via DCTcoefficient analysis,\textquotedblright}
{\it Pattern Recogn.},Vol. 42, pp. 2492-2250, 2009.

\bibitem{J20}
X. Pan, and S. Lyu, {\textquotedblleft Region duplication detection using image
feature matching,\textquotedblright}
{\it IEEE Trans. Inf. Foren. Sec.},Vol. 5, no. 4,
pp. 857-867, Dec. 2010.

\bibitem{J21}
R. C. Gonzalez, and R. E. Woods,, {\textquotedblleft Digital Image Processing
Using Matlab,\textquotedblright}
{\it Pearson Education India}, 2004.

\bibitem{J22}
Y. Huang, W. Lu, W. Sun, and D. Long, {\textquotedblleft Improved DCT-based
detection of copy-move forgery in images,\textquotedblright}
{\it Forensic Sci. Int.}, Vol. 206, pp. 178-184, 2011.

\bibitem{J23}
Yanjun Cao, T. Gao, and Qunting Yang, {\textquotedblleft A robust detection
algorithm for copy-move forgery in digital images,\textquotedblright}
{\it Forensic
Int.}, Vol. 214, pp. 33-43, 2012.


\bibitem{1}
G. Muhammad, M. Hussain, K. Khawaji and G. Bebis, {\textquotedblleft Blind copy move image forgery detection using dyadic undecimated wavelet transform,\textquotedblright}
{\it 17th International Conference on Digital Signal Processing (DSP)}, Corfu, 2011, pp. 1-6.

\bibitem{2}
N. Muhammad, M. Hussain, G. Muhammad and G. Bebis, {\textquotedblleft Copy-Move Forgery Detection Using Dyadic Wavelet Transform,\textquotedblright}
{\it Eighth International Conference on Computer Graphics, Imaging and Visualization (CGIV)}, Singapore, 2011, pp. 103-108.


\bibitem{3}
H. Yao, T. Qiao, Z. Tang, Y. Zhao and H. Mao, {\textquotedblleft Detecting Copy-Move Forgery Using Non-negative Matrix Factorization,\textquotedblright}
{\it Third International Conference on Multimedia Information Networking and Security},  Shanghai, 2011, pp. 591-594.

\bibitem{4}
P. Kakar and N. Sudha, {\textquotedblleft Detecting copy-paste forgeries using transform-invariant features,\textquotedblright}
{\it IEEE 15th International Symposium on Consumer Electronics (ISCE)},  Singapore, 2011, pp. 58-61.

\bibitem{5}
S. Murali, B. S. Anami and G. B. Chittapur, {\textquotedblleft Detection of Copy-Create Image Forgery Using Luminance Level Techniques,\textquotedblright}
{\it Third National Conference on Computer Vision, Pattern Recognition, Image Processing and Graphics (NCVPRIPG)}, Hubli, Karnataka, 2011, pp. 215-218.

\bibitem{6}
Dongmei Hou, Zhengyao Bai and Shuchun Liu, {\textquotedblleft Image copy-move forgery detection based on “crossing shadow” division,\textquotedblright}
{\it International Conference on Electric Information and Control Engineering (ICEICE)}, Wuhan, 2011,  pp. 1416-1419.

\bibitem{7}
B. Xu, G. Liu and Y. Dai, {\textquotedblleft A Fast Image Copy-Move Forgery Detection Method Using Phase Correlation,\textquotedblright}
{\it Fourth International Conference on Multimedia Information Networking and Security}, Nanjing, 2012, pp. 319-322.

\bibitem{8}
V. Christlein, C. Riess, J. Jordan, C. Riess and E. Angelopoulou, {\textquotedblleft An Evaluation of Popular Copy-Move Forgery Detection Approaches,\textquotedblright}
{\it IEEE Transactions on Information Forensics and Security}, vol. 7, no. 6, pp. 1841-1854, Dec. 2012.

\bibitem{9}
L. Li, S. Li and J. Wang, {\textquotedblleft Copy-move forgery detection based on PHT,\textquotedblright}
{\it World Congress on Information and Communication Technologies (WICT)}, pp. 1061-1065, Trivandrum, 2012.

\bibitem{10}
Xiaomei Quan and Hongbin Zhang, {\textquotedblleft Copy-move forgery detection in digital images based on local dimension estimation,\textquotedblright}
{\it International Conference on Cyber Security, Cyber Warfare and Digital Forensic (CyberSec)}, Kuala Lumpur, 2012, pp. 180-185.

\bibitem{11}
M. Hussain, G. Muhammad, S. Q. Saleh, A. M. Mirza and G. Bebis, {\textquotedblleft Copy-Move Image Forgery Detection Using Multi-Resolution Weber Descriptors,\textquotedblright}
{\it Eighth International Conference on Signal Image Technology and Internet Based Systems (SITIS)}, Naples, 2012, pp. 395-401.

\bibitem{12}
H. C. Hsu and M. S. Wang, {\textquotedblleft Detection of copy-move forgery image using Gabor descriptor,\textquotedblright}
{\it Anti-counterfeiting, Security, and Identification}, pp. 1-4, Taipei, 2012.

\bibitem{13}
H. C. Nguyen and S. Katzenbeisser, {\textquotedblleft Detection of Copy-move Forgery in Digital Images Using Radon Transformation and Phase Correlation,\textquotedblright}
{\it Eighth International Conference on Intelligent Information Hiding and Multimedia Signal Processing (IIH-MSP)}, Piraeus, 2012, pp. 134-137.

\bibitem{14}
Le Zhong and Weihong Xu, {\textquotedblleft A robust image copy-move forgery detection based on mixed moments,\textquotedblright}
{\it 4th IEEE International Conference on Software Engineering and Service Science (ICSESS)}, Beijing, 2013, pp. 381-384.

\bibitem{15}
S. Kumar, J. Desai and S. Mukherjee, {\textquotedblleft A fast DCT based method for copy move forgery detection,\textquotedblright}
{\it IEEE Second International Conference on Image Information Processing (ICIIP), 2013}, Shimla, 2013, pp. 649-654.

\bibitem{16}
M. F. Hashmi, A. R. Hambarde and A. G. Keskar, {\textquotedblleft Copy move forgery detection using DWT and SIFT features,\textquotedblright}
{\it 13th International Conference on Intellient Systems Design and Applications}, Bangi, 2013, pp. 188-193.

\bibitem{17}
G. Muhammad, M. H. Al-Hammadi, M. Hussain, A. M. Mirza and G. Bebis, {\textquotedblleft Copy move image forgery detection method using steerable pyramid transform and texture descriptor,\textquotedblright}
{\it IEEE EUROCON, 2013}, Zagreb, 2013, pp. 1586-1592.

\bibitem{18}
S. Chakraborty, {\textquotedblleft Copy move image forgery detection using mutual information,\textquotedblright}
{\it Fourth International Conference on Computing, Communications and Networking Technologies (ICCCNT),2013}, Tiruchengode, 2013, pp. 1-4.

\bibitem{19}
S. Ketenci and G. Ulutas, {\textquotedblleft Copy-move forgery detection in images via 2D-Fourier Transform,\textquotedblright}
{\it 36th International Conference on Telecommunications and Signal Processing (TSP), 2013}, Rome, 2013, pp. 813-816.

\bibitem{20}
M. AlSawadi, G. Muhammad, M. Hussain and G. Bebis, {\textquotedblleft Copy-Move Image Forgery Detection Using Local Binary Pattern and Neighborhood Clustering,\textquotedblright}
{\it Modelling Symposium (EMS), 2013}, pp. 249-254, European, Manchester, 2013.

\bibitem{21}
A. Kashyap and S. D. Joshi, {\textquotedblleft Detection of copy-move forgery using wavelet decomposition,\textquotedblright}
{\it International Conference on Signal Processing and Communication (ICSC), 2013}, Noida, 2013, pp. 396-400.

\bibitem{22}
M. B. Imamoglu, G. Ulutas and M. Ulutas, {\textquotedblleft Detection of copy-move forgery using Krawtchouk moment,\textquotedblright}
{\it 8th International Conference on Electrical and Electronics Engineering (ELECO), 2013}, Bursa, 2013, pp. 311-314.

\bibitem{23}
S. Y. Liao and T. Q. Huang, {\textquotedblleft Video copy-move forgery detection and localization based on Tamura texture features,\textquotedblright}
{\it 6th International Congress on Image and Signal Processing (CISP), 2013}, Hangzhou, 2013, pp. 864-868.

\bibitem{24}
M. F. Hashmi, V. Anand and A. G. Keskar, {\textquotedblleft A copy-move image forgery detection based on speeded up robust feature transform and Wavelet Transforms,\textquotedblright}
{\it International Conference on Computer and Communication Technology (ICCCT), 2014}, Allahabad, 2014, pp. 147-152.

\bibitem{25}
S. A. Fattah, M. M. I. Ullah, M. Ahmed, I. Ahmmed and C. Shahnaz, {\textquotedblleft A scheme for copy-move forgery detection in digital images based on 2D-DWT,\textquotedblright}
{\it IEEE 57th International Midwest Symposium on Circuits and Systems (MWSCAS)}, College Station, TX, 2014, pp. 801-804.

\bibitem{26}
M. Zandi, A. Mahmoudi-Aznaveh and A. Mansouri, {\textquotedblleft Adaptive matching for copy-move Forgery detection,\textquotedblright}
{\it IEEE International Workshop on Information Forensics and Security (WIFS)}, Atlanta, GA, pp. 119-124, 2014.

\bibitem{27}
D. Cozzolino, G. Poggi and L. Verdoliva, {\textquotedblleft Copy-move forgery detection based on PatchMatch,\textquotedblright}
{\it IEEE International Conference on Image Processing (ICIP)}, Paris, 2014, pp. 5312-5316.

\bibitem{28}
T. Chihaoui, S. Bourouis and K. Hamrouni, {\textquotedblleft Copy-move image forgery detection based on SIFT descriptors and SVD-matching,\textquotedblright}
{\it 1st International Conference on Advanced Technologies for Signal and Image Processing (ATSIP)}, Sousse, 2014, pp. 125-129.



\bibitem{29}
S. M. Fadl, N. A. Semary and M. M. Hadhoud, {\textquotedblleft Copy-rotate-move forgery detection based on spatial domain,\textquotedblright}
{\it 9th International Conference on Computer Engineering and Systems (ICCES)}, Cairo, 2014, pp. 136-141.

\bibitem{30}
L. Yu, Q. Han and X. Niu, {\textquotedblleft Copy-Rotation-Move Forgery Detection Using the MROGH Descriptor,\textquotedblright}
{\it IEEE International Conference on Cloud Engineering (IC2E), 2014}, Boston, MA, 2014, pp. 510-513.

\bibitem{31}
D. A. Ayalneh, H. J. Kim and Y. S. Choi, {\textquotedblleft ,\textquotedblright}
{\it 16th International Conference on Advanced Communication Technology}, Pyeongchang, 2014, pp. 181-185.

\bibitem{35}
Kashyap, Abhishek; B. Suresh; Agrawal, Megha; Gupta, Hariom; Joshi, Shiv Dutt, {\textquotedblleft Detection of splicing forgery using wavelet decomposition,\textquotedblright}
{\it IEEE International Conference on Computing, Communication and Automation (ICCCA)}, Noida, 15-16 May 2015, pp. 843-848.

\bibitem{32}
K. Sudhakar, V. M. Sandeep and S. Kulkarni, {\textquotedblleft Shape Based Copy Move Forgery Detection Using Level Set Approach,\textquotedblright}
{\it Fifth International Conference on Signal and Image Processing (ICSIP), 2014}, Jeju Island, 2014, pp. 213-217.

\bibitem{33}
K. Sudhakar, V. M. Sandeep and S. Kulkarni, {\textquotedblleft Speeding-up SIFT based copy move forgery detection using level set approach,\textquotedblright}
{\it International Conference on Advances in Electronics, Computers and Communications (ICAECC), 2014}, Bangalore, 2014, pp. 1-6.

\bibitem{34}
D. Tralic, S. Grgic and B. Zovko-Cihlar, {\textquotedblleft Video frame copy-move forgery detection based on Cellular Automata and Local Binary Patterns,\textquotedblright}
{\it 10th International Symposium on Telecommunications (BIHTEL), 2014}, Sarajevo, 2014, pp. 1-4.





%
%




%
%
%
%
%
%
%
%
%
%
%
%
%
%
%
%
%
%
%
%
%
%
%
%
%
%



%
%
%
%
%
%


\end{thebibliography}

\end{document}